\documentclass[%
 reprint,
superscriptaddress,
 amsmath,amssymb,
 aps,showkeys,prl,
]{revtex4-1}

\usepackage{float}
\usepackage{amsmath}
\usepackage{booktabs}
\usepackage{graphicx}
\usepackage{dcolumn}
\usepackage{bm}
\usepackage{xcolor,soul}
\usepackage{subfigure}
\usepackage{transparent}
\usepackage{soul}
\usepackage[normalem]{ulem}
\usepackage{braket}

\usepackage{graphicx,color,xcolor,colortbl}
\usepackage{hyperref}
\definecolor{darkgreen}{RGB}{0, 100, 0}

\newcommand{\e}[1]{\text{#1}}

\begin{document}

\preprint{APS/123-QED}

\title{Multiphoton dressed Rydberg excitations in a microwave cavity\\ with ultracold Rb atoms} 

\author{J.~D.~Massayuki Kondo}
\affiliation{Departamento de Física, Universidade Federal de Santa Catarina, Florianópolis 88040-900, SC, Brasil}
\affiliation{Instituto de F\'{i}sica de S\~{a}o Carlos, Universidade de S\~{a}o Paulo, Caixa Postal 369, 13560-970, S\~{a}o Carlos, SP, Brasil}
\author{Seth T.~Rittenhouse}
\affiliation{Department of Physics, the United States Naval Academy, Annapolis, Maryland 21402, USA}
\affiliation{ITAMP, Center for Astrophysics $|$ Harvard \& Smithsonian Cambridge, Massachusetts 02138, USA}
\affiliation{Institute of Theoretical Physics, Institute of Physics, University of Amsterdam, Science Park 904, 1098 XH Amsterdam, The Netherlands}
\author{Daniel Varela Magalhães}
\affiliation{Instituto de F\'{i}sica de S\~{a}o Carlos, Universidade de S\~{a}o Paulo, Caixa Postal 369, 13560-970, S\~{a}o Carlos, SP, Brasil}
\author{Vasil Rokaj}
\affiliation{ITAMP, Center for Astrophysics $|$ Harvard \& Smithsonian Cambridge, Massachusetts 02138, USA}
\affiliation{Department of Physics, Harvard University, Cambridge, USA}
\author{S.~I.~Mistakidis}
\affiliation{ITAMP, Center for Astrophysics $|$ Harvard \& Smithsonian Cambridge, Massachusetts 02138, USA}
\affiliation{Department of Physics, Missouri University of Science and Technology, Rolla, MO 65409, USA}
\author{H.~R.~Sadeghpour}
\affiliation{ITAMP, Center for Astrophysics $|$ Harvard \& Smithsonian Cambridge, Massachusetts 02138, USA}
\author{Luis Gustavo Marcassa}
\affiliation{Instituto de F\'{i}sica de S\~{a}o Carlos, Universidade de S\~{a}o Paulo, Caixa Postal 369, 13560-970, S\~{a}o Carlos, SP, Brasil}

\date{\today}

\begin{abstract}
We investigate magneto-optical trap loss spectroscopy of Rydberg excited $^{85}$Rb ($66\leq n \leq 68~S_{1/2}$) atoms, placed inside a tailored microwave cavity. The cavity frequency at 13.053 GHz is in resonance with the $67S_{1/2} \rightarrow 66P_{3/2}$ transition, inducing a ladder multiphoton microwave Rydberg absorption and emission. The observed spectra are modeled with an extended Jaynes-Cumming formalism that accounts for multiphoton absorption from and emission into the cavity, the loss from the trap due to Rydberg excitation, and cavity imperfection. We calculate the average photons in each spectral feature and find evidence for fractional photon emission into the cavity modes. The microwave cavity Rydberg spectroscopy in this work should inform the application and technology development of Rydberg based sensors and hybrid Rydberg atom-superconducting resonator quantum gates.
\end{abstract}

                   
\maketitle

Rydberg excitation in a cavity has a celebrated history in the emergence of cavity Quantum Electrodynamics (cQED). The Purcell effect~\cite{Purcell1946} occurs when an increase in the local density of microwave (MW) photons leads to enhanced spontaneous emission. Initial observations were made on sodium atomic beams by enhancing \cite{Haroche1983,HAROCHE1985,haroche1989cavity,haroche1999cavity} and suppressing \cite{Klepnner1981,Hulet1985} spontaneous emission. Additional experiments observed deviations in the Rydberg state lifetime in ultracold samples due to geometrical dependencies of the trapping vacuum chamber \cite{Archimi2019,PhysRevA.82.063406,Magnani2020}, showing a direct relation between black body radiation spectra and allowed cavity modes \cite{PhysRevA.87.033801,PhysRevA.100.030501,PhysRevA.92.012517}.

A well-tuned MW cavity can be used as a diagnostic tool to characterize cold plasma densities and temperatures~\cite{ninhuijs2021design}.
More recently, cold Rydberg atoms interacting with MW cavity photons have been proposed to build a MW-to-optical converter via four-wave and six-wave mixing \cite{Covey2019,PhysRevLett.120.093201} and create quantum transducers from millimeter wave to optical field in a hybrid superconductor cryogenic resonator with ultracold Rb atoms \cite{Kumar2023Quantum,PhysRevA.108.022805}. Advances in Rydberg field sensors \cite{Wade2017,Wade2018} and Rydberg based logic gates \cite{Sebadi2022,Jau2016} require a deep understanding of the intricacies of MW dressing of Rydberg states. MW  resonators and cavities have been also applied in combination with Electromagnetically Induced Transparency  (EIT) to enhance MW electrometry in thermal samples \cite{Hollowy2022,liu2024,simons2019embedding}. 

We use an ultracold sample of $^{85}\text{Rb}$ held in a magneto-optical trap (MOT) placed inside a MW cavity to excite $nS_{1/2}$ Rydberg states ($66\leq n \leq 68$). An extended Jaynes-Cumming model based on a multilevel multi-MW-photon formalism captures the details of the interaction of the MW cavity with the Rydberg atom excitation and identifies the multiphoton processes in the Rydberg fluorescence loss spectra. The model accounts for the loss of atoms from the cavity and faithfully reproduces, without any free parameters, the main features of the observed spectra. We calculate the average photon number for each spectral peak and find evidence for fractional photon emission into the cavity modes.

\begin{figure}[ht]
\centering
\includegraphics[scale=0.5]{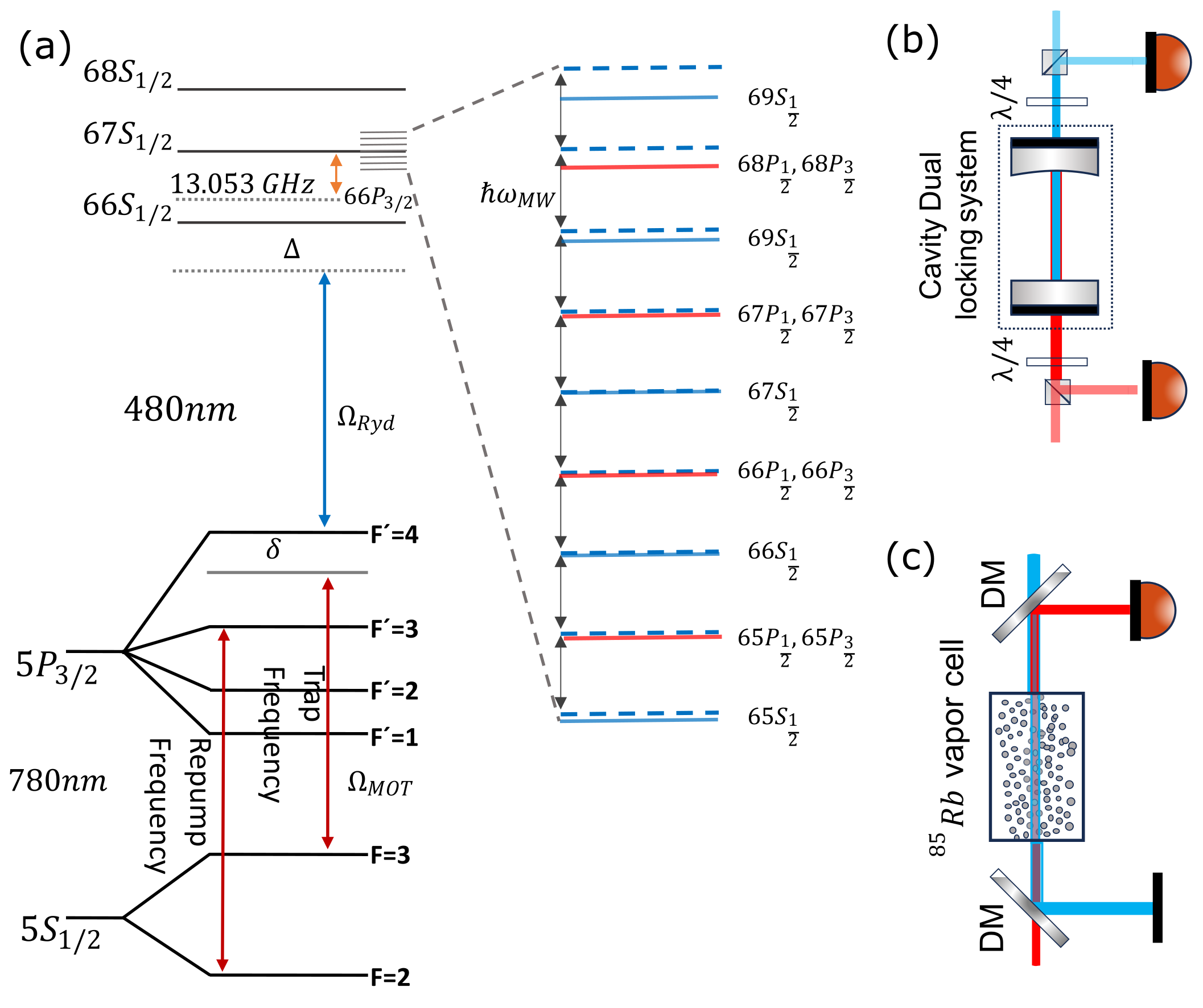}
\caption{\label{fig:fig1}(a) The ${}^{85}\text{Rb}$ energy diagram is shown with an enlarged view of the Rydberg spectrum near the 67$S_{1/2}$ level. The total amount of excitations (matter plus MW photon) in the system is conserved. The dressed Rydberg basis set includes excitation from 
$\ket{67S_{1/2},n_0}$ to
$\dots 
\left\vert66P_{J},n_{0}+1\right\rangle, \left\vert67S_{1/2},n_{0}\right\rangle, \left\vert67P_{J},n_{0}-1\right\rangle,\dots$
The number of cavity photons in the system is extracted by setting the Rabi frequency of the $67S_{1/2}\rightarrow66P_{3/2}$ transition, equal to the splitting between the two most prominent features in the experimental spectrum.  The number of states included is controlled by the number of photons emitted or absorbed by the atom, $\Delta n$.  We find that the AC Stark energies for excitations near the 67S$_{1/2}$ are converged by including all states with $|\Delta n| \le 4$.  (b) Schematic representation of the EIT vapour reference cell, and (c) the thermally stabilized Fabry-Perot optical cavity (dual-locking system).}
\end{figure} 

The $^{85}$Rb MOT is loaded from an atomic vapor cell at room temperature and operates in a stainless steel chamber with a background pressure below $10^{-9}$ torr. Under normal conditions, it traps approximately $10^{7}$ atoms at a density of about $10^{10}$ cm$^{-3}$. The trapping laser beam is red-tuned from the $5S_{1/2},F=3\rightarrow5P_{3/2},F'=4$ atomic transition, with an average Rabi frequency $\Omega_{MOT}/2\pi$ of $13$ MHz, detuned by $\delta \approx-2.1\Gamma$, where $\Gamma/2\pi=5.9$ MHz. Fig. \ref{fig:fig1}(a) shows the ${}^{85}$Rb energy states involved in trapping, cooling and excitation of a $nS$ Rydberg state. The repumping laser beam is resonant with the $5S_{1/2},F=2\rightarrow5P_{3/2},F'=3$ atomic transition. The trapping light is frequency-locked to a thermally stabilized optical cavity shown in Fig. $\ref{fig:fig1}$(b) \cite{RodriguezFernandez2023}, and the repumping laser frequency is stabilized using a compact saturation spectroscopy system. Another laser, operating at a wavelength of 480 nm with a $300~\mu$m waist and $4~\e{W}/\e{cm}^2$ intensity, is used to couple the $5P_{3/2},F'=4$ state with the ${nS_{1/2}}$ Rydberg states in the range $66\leq n \leq 68$. 

The targeted Rydberg states are motivated because the cavity is resonant with the $67S_{1/2} \rightarrow 66P_{3/2}$ transition, allowing the observation of multiphoton MW Rydberg transitions. This laser is frequency-locked to the same optical cavity after modulation by an electro-optical modulator, enabling precise scanning of the blue Rydberg laser in frequency steps as short as 100 kHz. The frequency of the Rydberg laser can be continuously monitored and calibrated using an EIT signal from a reference vapor cell held at ambient temperature (see Fig. $\ref{fig:fig1}$(c)). The 780 nm probe beam is provided by the trapping laser through the use of an acoustic-optical modulator (AOM), such that its detuning is set to zero ($\delta=0$) with respect to the trapping transition. As the 480 nm laser scans over the EIT $nS_{1/2}$ resonances, its zero detuning ($\Delta=0$) is set at the $5P_{3/2},F=4 \rightarrow nS_{1/2}$ transition. 

\begin{figure}[ht]
\centering
\includegraphics[scale=0.37]{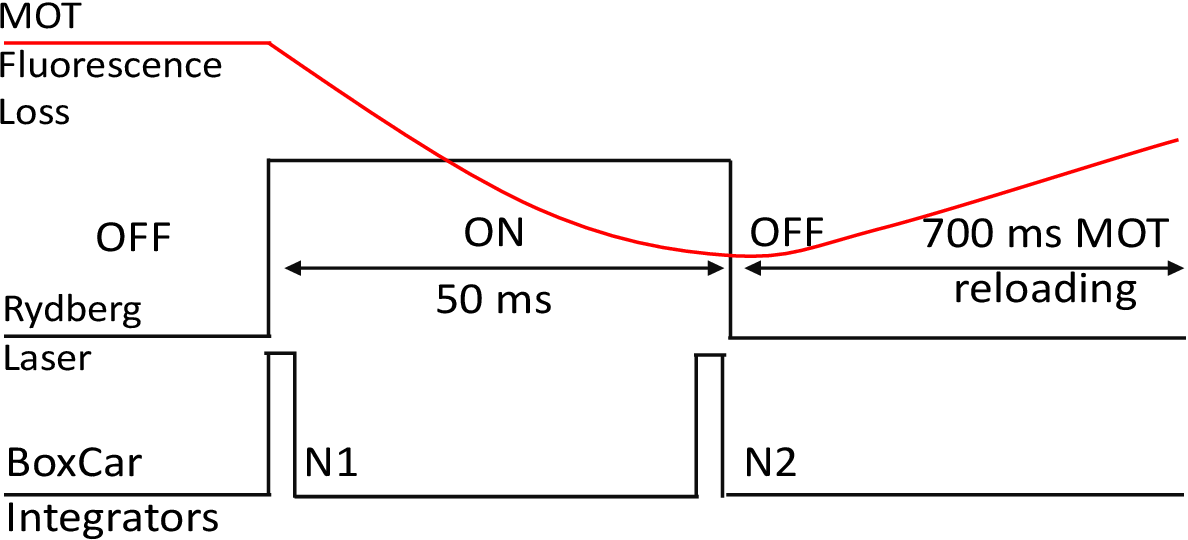}
\caption{\label{fig:fig3}Schematic of the experimental time sequence. The MOT fluorescence signal is acquired via boxcar integration electronics.}
\end{figure}

As for a specific resonant blue laser detuning $\Delta$, a number of Rydberg excitations can dynamically depopulate the atomic trap while background ground state cold atoms are replenishing it in a constant rate. Toggling the blue laser during 50 milliseconds proved to be sufficient for measuring the differences in MOT fluorescence signal and allow it to reach a steady-state population. The time sequence implemented is presented in Fig. \ref{fig:fig3}. The fluorescence signal is acquired by a fast photodiode and fed to a boxcar rapid integrator, using two integration time windows of $15~\mu$s each. Firstly, an N1 integration is performed just after turning the Rydberg laser on, and secondly, an N2 integration is carried just before turning it off, with the MOT reloaded for 700 milliseconds before next iteration. The ratio N2/N1 was recorded for every $\Delta$ and accounted for the eventual MOT population fluctuations with an average of 4 measurements per experimental point.

The full level scheme of the system is illustrated in Fig.~\ref{fig:fig1}(a). The $13.053$ GHz MW cavity mode is resonant with the $67S_{1/2} \leftrightarrow 66P_{3/2}$ Rydberg transition. The detailed experimental setup is described in the Supplemental Material \cite{supp}. Moreover, the $nS_{1/2}$ state is nearly halfway between the $\left(  n-1\right)  P_{J}$ and the $nP_{J}$ states. This implies that the cavity is almost resonant with the $67S_{1/2} \leftrightarrow 67P_{3/2}$ transition, and thus can also drive the transition to the $67P_{3/2}$ state. The small spin-orbit splitting at high principal quantum number \cite{LiPRA2003} necessitates the inclusion of the $nP_{1/2}$ states in our theoretical model. Because the Rydberg states are not trapped by the MOT, they are observed  as a loss signal from the cold atomic gas which we shall model as a non-hermitian loss. As a result the corresponding Hamiltonian for the  Rydberg atom excitation in the MW cavity is described by an extended (multilevel), non-Hermitian Jaynes-Cummings model,
\begin{equation}
\hat{H}_{MW}=\hbar\omega_{MW}\hat{a}^{\dag}\hat{a}+\frac{\hbar}{2}\left(
\hat{\Sigma}^{+}\hat{a}+\hat{\Sigma}^{-}\hat{a}^{\dag}\right)  +\hat{H}_{Ryd}.
\label{Eq:H_MW}%
\end{equation} 
The operators $\hat{a}$ ($\hat{a}^{\dag}$) annihilate (create) cavity photons, and $\hat{H}_{Ryd}$ is the uncoupled Rydberg Hamiltonian:
\begin{equation}
    \hat{H}_{Ryd}=\sum_{nLJ}\left(-\frac{E_{Ryd}}{\left(  n-\mu_{LJ}\right)  ^{2}}-i\hbar\frac{\gamma
}{2}\right)\left|nLJ\right>\left<nLJ\right|. \label{Eq:HRydprime}
\end{equation} 
Here, $\mu_{LJ}$ denotes the Rydberg quantum defect \cite{LiPRA2003} for orbital and total electron angular momenta ($L$ and $J$),  $E_{Ryd}$ is the Rydberg constant, and $\gamma$ represents the phenomenological loss rate of Rydberg atoms from the trap.
In  Eq.~(\ref{Eq:H_MW}), $\hat{\Sigma}^{+}$=$\left(  \hat{\Sigma}^{-}\right)
^{\dag}$ is the operator representing the excitation of a Rydberg state through
the absorption (emission) of a photon. Since the $nD_{J}$ states are far detuned
from the cavity, we only consider the $nP_{J}$ and $nS_{1/2}$
states here. With this restriction the excitation operator reads
\begin{align}
\hat{\Sigma}^{+}=\sum_{n,J}&\left(  \Omega_{J}^{(n+1,n)}\left\vert \left(
n+1\right)  S_{1/2}\right\rangle \left\langle nP_{J}\right\vert \right.  \label{Eq:Rabi_Mat}\\
& \left. +\Omega_{J}^{(n,n)}\left\vert nP_{J}\right\rangle \left\langle nS_{1/2}\right\vert\right)  \nonumber,%
\end{align}
where $\Omega_{J}^{\left(  n^{\prime},n\right)  }=\sqrt
{\frac{2\hbar\omega_{MW}}{\varepsilon_{0}V}}\left\langle n^{\prime}%
S_{1/2}\left\vert \hat{d}\right\vert nP_{J}\right\rangle $ is the single photon Rabi
frequency,
$\varepsilon_{0}$ is the vacuum permittivity, and $V$ is the effective volume of the cavity mode. Because $\Omega_{J}^{\left(  n+1,n\right)
}$ varies by less than 10\% for  $\Delta n\approx3$, we treat $\Omega_{J}^{\left(  n^{\prime
},n\right)  }$ to be independent of $n$ over the range of Rydberg states of
interest, i.e. $\Omega_{J}^{\left(  n+1,n\right)  }\approx\Omega
_{J}^{\left(  n,n\right)  }\approx\Omega_{J}^{\left(  67,66\right)  }=$
$\Omega_{J}$.  Incorporating the two-photon excitation from the $\left|5S_{1/2}\right>$ to the $\left|nS_{1/2}\right>$ Rydberg state, we predict the atom loss rate from the MOT \cite{supp}. 

The final fraction of atoms remaining in the trap after a long Rydberg pulse reads 
\[
\frac{N}{N_0} = \sqrt{1+\left(\frac{\bar{\Gamma}_0/\gamma}{2 \eta}\right)^2}-\frac{\bar{\Gamma}_0/\gamma}{2\eta},
\]
where $N_{0}$ is the number of trapped atoms at the start of the pulse, $\bar{\Gamma}_0$ is the rate at which atoms are lost from the trap averaged over a Lorentzian cavity line profile, and $\eta=0.0125$ is a fitting parameter which can be thought of as approximately the fraction of atoms that would remain if the full Rydberg loss rate, $\bar{\Gamma}_0=\gamma$, were to be achieved.

\begin{figure}[ht]
\centering
\includegraphics[width=3in]{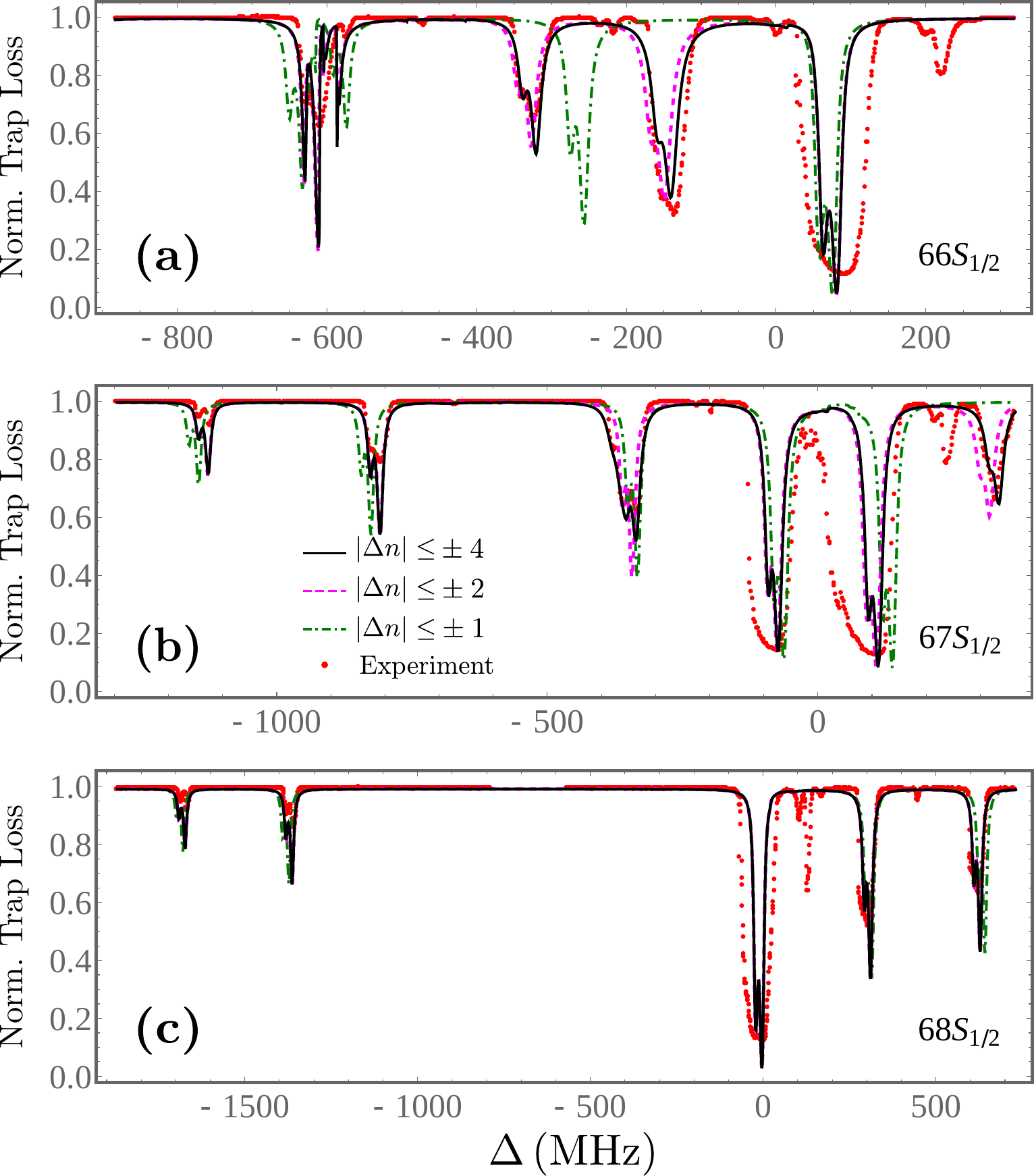} 
\caption{Normalized trap loss rate obtained from the extended Jaynes-Cummings model accounting for different  photon excitations (see legend) for varying Rydberg laser detuning, $\Delta$; for (a) 66$S_{1/2}$, (b) 67$S_{1/2}$ and (c) 68$S_{1/2}$. 
Each resonance has a double peak structure with a separation of $\sim 13$ MHz which originates from the AT splitting of the MOT laser coupling the $5S_{1/2}$ and $5P_{3/2}$ states \cite{Wang2023AT}. 
In all cases, the respective experimental curves are provided showing good agreement with the theoretical predictions.}
\label{fig:fig8}
\end{figure}

\begin{figure}[ht]
\centering
\includegraphics[width=3.25in]{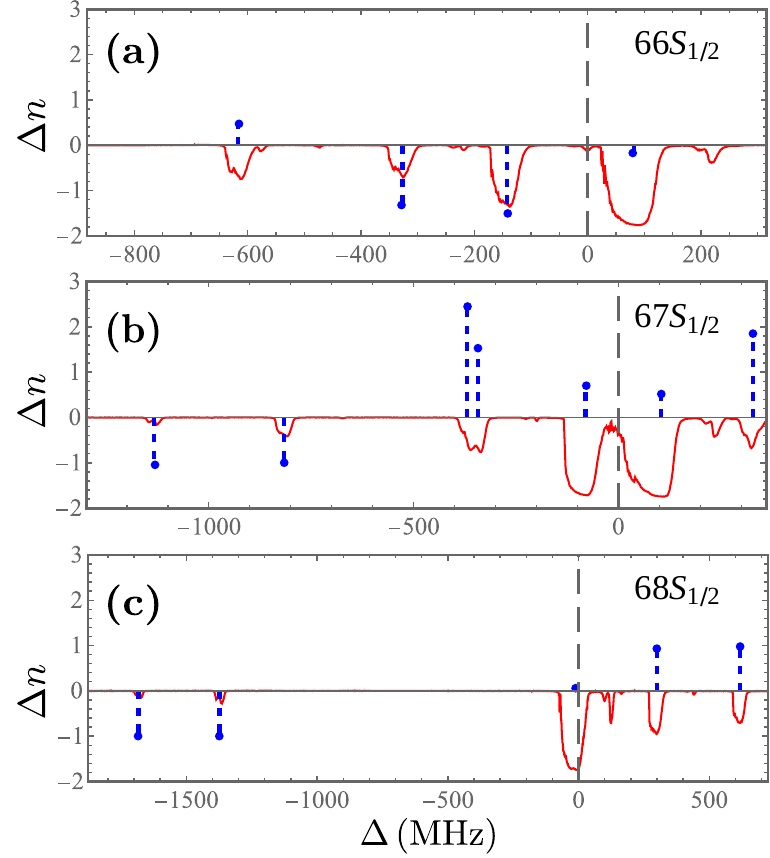} 
 
\caption{Average number of photons emitted into ($\Delta n >0$) or absorbed from ($\Delta n <0$) the cavity for each state. The cases of the (a) $66S_{1/2}$, (b) $67S_{1/2}$ and (c) $68S_{1/2}$ states are shown with respect to the Rydberg laser detuning, $\Delta$. The MW cavity is driven at 32.8$\%$ of the maximum driving power resulting in a MW electric field amplitude of 77.5 mV/cm. For reference, the experimental spectrum is depicted as a thin red curve.}
\label{fig:67S}
\end{figure}

Fig.~\ref{fig:fig8} illustrates a comparison of the observed trap loss spectra for different initial Rydberg states, $66S_{1/2}$, $67S_{1/2}$, and $68S_{1/2}$, with different levels of theory. The latter contains processes involving one, two and more photons, at a cavity field amplitude of 77.5 mV/cm. 
For each Rydberg state, there is a hyperfine state observed to the blue of the central Rydberg line that is not included in the theoretical description. It can be readily seen that many of the features for the $67S_{1/2}$ excitation spectrum depicted in Fig.~\ref{fig:fig8}(b) are approximately described by the inclusion of single photon transitions (i.e. only accounting for the $66P_{1/2,3/2}$ and $67P_{1/2,3/2}$ states in the Rydberg basis beyond the $67S_{1/2}$ state). 
Importantly however, in the spectra presented in  Fig.~\ref{fig:fig8} we observe phenomena which require the participation of higher-photon processes. 
The resonance appearing at a detuning of $+328$ MHz is only present in the theoretical curves which include two or more photon processes.  There is also a state at a detuning of $-370$ MHz (apparent as an additional shoulder to the red of the line at $-350$ MHz) in Fig.~\ref{fig:fig8}(b) which is only accounted for when three or more photon processes are included. 
Notice that the location of all states in the spectrum is numerically converged only when including higher multiphoton processes.

In Fig.~\ref{fig:67S}, we include the specific MW transitions showing the average number of photons emitted into ($\Delta n > 0$) or absorbed from ($\Delta n<0$) the cavity for each Rydberg line in the $66S_{1/2}$, $67S_{1/2}$, and $68S_{1/2}$ spectra at the cavity field amplitude of 77.5 mV/cm. 
The two resonances occurring at -142 and -327 MHz in Fig.~\ref{fig:67S}(a) refer to AT splitting and are shifted from the main AT lines in  Fig.~\ref{fig:67S}(b), by absorption of an additional photon from the cavity in the multiphoton process $66S_{1/2}\rightarrow 66P_{3/2}\rightarrow 67S_{1/2}$. The line at -631 MHz is a single photon absorption to the $65P_{3/2}$ state, and the main Rydberg line is AC Stark blue-shifted. In Fig.~\ref{fig:67S}(b), the resonances at +328 MHz and the one at $-370$ MHz, are 2- and 3-photon processes associated with the $\left\vert 67S_{1/2},n_{0}\right\rangle \rightarrow \left\vert 66P_{3/2},n_{0}+1\right\rangle\rightarrow \left\vert 66S_{1/2},n_{0}+2\right\rangle$
and and additional $\left\vert 66S_{1/2},n_{0}+2\right\rangle\rightarrow \left\vert 65P_{3/2},n_{0}+3\right\rangle$
transitions, respectively.   The resonance at -342 MHz is largely a single photon transition, $\left\vert 67S_{1/2},n_{0}\right\rangle \rightarrow \left\vert 66P_{1/2},n_{0}+1\right\rangle$, with some additional coupling to the $\left\vert 66S_{1/2},n_{0}+2\right\rangle$ state.  The resonances in the detuning range -1100 MHz to -800 MHz are single-photon emission lines $\left\vert 67S_{1/2},n_{0}\right\rangle \rightarrow \left\vert 67P_{3/2},n_{0}+1\right\rangle$ and $\left\vert 67S_{1/2},n_{0}\right\rangle \rightarrow \left\vert 67P_{1/2},n_{0}+1\right\rangle $, respectively. 
Because the cavity is far detuned from any $nP$ states near the $68S_{1/2}$ state, see Fig.~\ref{fig:fig1}(a), only single-photon emission into and absorption from the MW cavity are visible in Fig.~\ref{fig:67S}(c).

An interesting feature  in Fig.~\ref{fig:67S}(b) is that there exist peaks which share one photon. This is true for the doublets in Fig.~\ref{fig:67S}(b) around zero detuning and around -370 MHz. This phenomenon indicates the existence of non-trivial photon correlations between the states that share the photon excitation. To inspect the type of photon correlations, we employ the standard two-level Jaynes-Cummings model. The dressed Jaynes-Cummings states for zero detuning are $|\pm, n_0\rangle =\frac{1}{\sqrt{2}}\left(|g,n_0+1\rangle \pm |e,n_0\rangle \right)$ where $g$ and $e$ are the lowest and highest state respectively in the two-level system~\cite{Gerry_Knight_2004}, and $n_0$ the number of photons. The dressed states have non-integer photon numbers $\langle \hat{a}^{\dagger}\hat{a}\rangle_{\pm n_0}=n_0+\frac{1}{2}$ and they share a photon. 

To explore the nature of photon correlations, we resort to the Mandel Q parameter~\cite{Gerry_Knight_2004},  
$$Q=\frac{\langle \hat{a}^{\dagger} \hat{a}^{\dagger} \hat{a}\hat{a}\rangle -\langle \hat{a}^{\dagger} \hat{a}\rangle^2}{\langle \hat{a}^{\dagger} \hat{a}\rangle}$$, 
which in our case is found to be negative, i.e. $Q_{\pm n_0}=-\frac{n_0+\frac{1}{4}}{n_0+\frac{1}{2}}$.  
This analysis suggests the possibility that the spectral peaks in Fig.~\ref{fig:67S}(b) featuring non-integer photon numbers exhibit non-classical photon correlations.

We explored the direct excitation of dressed Rydberg states in an ultra-cold ${}^{85}\text{Rb}$ sample held inside a three dimensional MW cavity and observed multiphotonic MW cavity exchange. For this we investigate the excitation spectra around three specific Rydberg states $66\e{S}_{1/2}$, $67\e{S}_{1/2}$ and $68\e{S}_{1/2}$ by performing fluorescence trap-loss spectroscopy. The cavity was engineered to match the energy difference of the transition $67\e{S}_{1/2}\rightarrow 66\e{P}_{3/2}$ and MW field was supplied to the chamber by an intracavity antenna. We were able to describe the Rydberg excitation spectra by using a multilevel Jaynes-Cumming model. By taking into account multiphotonic processes of emission and absorption within the MW resonant cavity, and including a precise trap loss description, the model greatly reproduces the observed spectra without any free parameters. We also calculate the average number of MW photons involved for all the main spectral features and identify the resonances with emission into and absorption from cavity MW photons. It would be interesting to probe quantum effects due to the cavity for the generation of specific dressed states, as superposition of photon excitations. The fractional MW photon numbers in the Rydberg AT lines in Fig.~\ref{fig:67S}(b) may herald the creation of non-classical photon states and photon anti-bunching. Further investigations of the MW photon statistics by cooling the cavity are warranted. 

\begin{acknowledgments}
This work is supported by grants 2019/10971-0 and 2021/06371-7, S\~{a}o Paulo Research Foundation (FAPESP), and CNPq (305257/2022-6). It was also supported by the Army Research Office - Grant Number W911NF-21-1-0211. V.R., S.I.M, H.R.S.~and S.T.R.~acknowledge support from the NSF through a grant for ITAMP at Harvard University.
\end{acknowledgments}

\bibliography{References}

\providecommand{\noopsort}[1]{}\providecommand{\singleletter}[1]{#1}%
\begin{thebibliography}{30}%
\makeatletter
\providecommand \@ifxundefined [1]{%
 \@ifx{#1\undefined}
}%
\providecommand \@ifnum [1]{%
 \ifnum #1\expandafter \@firstoftwo
 \else \expandafter \@secondoftwo
 \fi
}%
\providecommand \@ifx [1]{%
 \ifx #1\expandafter \@firstoftwo
 \else \expandafter \@secondoftwo
 \fi
}%
\providecommand \natexlab [1]{#1}%
\providecommand \enquote  [1]{``#1''}%
\providecommand \bibnamefont  [1]{#1}%
\providecommand \bibfnamefont [1]{#1}%
\providecommand \citenamefont [1]{#1}%
\providecommand \href@noop [0]{\@secondoftwo}%
\providecommand \href [0]{\begingroup \@sanitize@url \@href}%
\providecommand \@href[1]{\@@startlink{#1}\@@href}%
\providecommand \@@href[1]{\endgroup#1\@@endlink}%
\providecommand \@sanitize@url [0]{\catcode `\\12\catcode `\$12\catcode
  `\&12\catcode `\#12\catcode `\^12\catcode `\_12\catcode `\%12\relax}%
\providecommand \@@startlink[1]{}%
\providecommand \@@endlink[0]{}%
\providecommand \url  [0]{\begingroup\@sanitize@url \@url }%
\providecommand \@url [1]{\endgroup\@href {#1}{\urlprefix }}%
\providecommand \urlprefix  [0]{URL }%
\providecommand \Eprint [0]{\href }%
\providecommand \doibase [0]{http://dx.doi.org/}%
\providecommand \selectlanguage [0]{\@gobble}%
\providecommand \bibinfo  [0]{\@secondoftwo}%
\providecommand \bibfield  [0]{\@secondoftwo}%
\providecommand \translation [1]{[#1]}%
\providecommand \BibitemOpen [0]{}%
\providecommand \bibitemStop [0]{}%
\providecommand \bibitemNoStop [0]{.\EOS\space}%
\providecommand \EOS [0]{\spacefactor3000\relax}%
\providecommand \BibitemShut  [1]{\csname bibitem#1\endcsname}%
\let\auto@bib@innerbib\@empty
\bibitem [{\citenamefont {Purcell}\ \emph {et~al.}(1946)\citenamefont
  {Purcell}, \citenamefont {Torrey},\ and\ \citenamefont
  {Pound}}]{Purcell1946}%
  \BibitemOpen
  \bibfield  {author} {\bibinfo {author} {\bibfnamefont {E.~M.}\ \bibnamefont
  {Purcell}}, \bibinfo {author} {\bibfnamefont {H.~C.}\ \bibnamefont {Torrey}},
  \ and\ \bibinfo {author} {\bibfnamefont {R.~V.}\ \bibnamefont {Pound}},\
  }\href {\doibase 10.1103/PhysRev.69.37} {\bibfield  {journal} {\bibinfo
  {journal} {Phys. Rev.}\ }\textbf {\bibinfo {volume} {69}},\ \bibinfo {pages}
  {37} (\bibinfo {year} {1946})}\BibitemShut {NoStop}%
\bibitem [{\citenamefont {Goy}\ \emph {et~al.}(1983)\citenamefont {Goy},
  \citenamefont {Raimond}, \citenamefont {Gross},\ and\ \citenamefont
  {Haroche}}]{Haroche1983}%
  \BibitemOpen
  \bibfield  {author} {\bibinfo {author} {\bibfnamefont {P.}~\bibnamefont
  {Goy}}, \bibinfo {author} {\bibfnamefont {J.~M.}\ \bibnamefont {Raimond}},
  \bibinfo {author} {\bibfnamefont {M.}~\bibnamefont {Gross}}, \ and\ \bibinfo
  {author} {\bibfnamefont {S.}~\bibnamefont {Haroche}},\ }\href {\doibase
  10.1103/PhysRevLett.50.1903} {\bibfield  {journal} {\bibinfo  {journal}
  {Phys. Rev. Lett.}\ }\textbf {\bibinfo {volume} {50}},\ \bibinfo {pages}
  {1903} (\bibinfo {year} {1983})}\BibitemShut {NoStop}%
\bibitem [{\citenamefont {Haroche}\ and\ \citenamefont
  {Raimond}(1985)}]{HAROCHE1985}%
  \BibitemOpen
  \bibfield  {author} {\bibinfo {author} {\bibfnamefont {S.}~\bibnamefont
  {Haroche}}\ and\ \bibinfo {author} {\bibfnamefont {J.}~\bibnamefont
  {Raimond}}\ }(\bibinfo  {publisher} {Academic Press},\ \bibinfo {year}
  {1985})\ pp.\ \bibinfo {pages} {347--411}\BibitemShut {NoStop}%
\bibitem [{\citenamefont {Haroche}\ and\ \citenamefont
  {Kleppner}(1989)}]{haroche1989cavity}%
  \BibitemOpen
  \bibfield  {author} {\bibinfo {author} {\bibfnamefont {S.}~\bibnamefont
  {Haroche}}\ and\ \bibinfo {author} {\bibfnamefont {D.}~\bibnamefont
  {Kleppner}},\ }\href {\doibase 10.1063/1.881201} {\bibfield  {journal}
  {\bibinfo  {journal} {Physics Today}\ }\textbf {\bibinfo {volume} {42}},\
  \bibinfo {pages} {24} (\bibinfo {year} {1989})}\BibitemShut {NoStop}%
\bibitem [{\citenamefont {Haroche}(1999)}]{haroche1999cavity}%
  \BibitemOpen
  \bibfield  {author} {\bibinfo {author} {\bibfnamefont {S.}~\bibnamefont
  {Haroche}},\ }\href {\doibase 10.1063/1.58235} {\bibfield  {journal}
  {\bibinfo  {journal} {AIP Conference Proceedings}\ }\textbf {\bibinfo
  {volume} {464}},\ \bibinfo {pages} {45} (\bibinfo {year} {1999})}\BibitemShut
  {NoStop}%
\bibitem [{\citenamefont {Kleppner}(1981)}]{Klepnner1981}%
  \BibitemOpen
  \bibfield  {author} {\bibinfo {author} {\bibfnamefont {D.}~\bibnamefont
  {Kleppner}},\ }\href {\doibase 10.1103/PhysRevLett.47.233} {\bibfield
  {journal} {\bibinfo  {journal} {Phys. Rev. Lett.}\ }\textbf {\bibinfo
  {volume} {47}},\ \bibinfo {pages} {233} (\bibinfo {year} {1981})}\BibitemShut
  {NoStop}%
\bibitem [{\citenamefont {Hulet}\ \emph {et~al.}(1985)\citenamefont {Hulet},
  \citenamefont {Hilfer},\ and\ \citenamefont {Kleppner}}]{Hulet1985}%
  \BibitemOpen
  \bibfield  {author} {\bibinfo {author} {\bibfnamefont {R.~G.}\ \bibnamefont
  {Hulet}}, \bibinfo {author} {\bibfnamefont {E.~S.}\ \bibnamefont {Hilfer}}, \
  and\ \bibinfo {author} {\bibfnamefont {D.}~\bibnamefont {Kleppner}},\ }\href
  {\doibase 10.1103/PhysRevLett.55.2137} {\bibfield  {journal} {\bibinfo
  {journal} {Phys. Rev. Lett.}\ }\textbf {\bibinfo {volume} {55}},\ \bibinfo
  {pages} {2137} (\bibinfo {year} {1985})}\BibitemShut {NoStop}%
\bibitem [{\citenamefont {Archimi}\ \emph
  {et~al.}(2019{\natexlab{a}})\citenamefont {Archimi}, \citenamefont
  {Simonelli}, \citenamefont {Di~Virgilio}, \citenamefont {Greco},
  \citenamefont {Ceccanti}, \citenamefont {Arimondo}, \citenamefont {Ciampini},
  \citenamefont {Ryabtsev}, \citenamefont {Beterov},\ and\ \citenamefont
  {Morsch}}]{Archimi2019}%
  \BibitemOpen
  \bibfield  {author} {\bibinfo {author} {\bibfnamefont {M.}~\bibnamefont
  {Archimi}}, \bibinfo {author} {\bibfnamefont {C.}~\bibnamefont {Simonelli}},
  \bibinfo {author} {\bibfnamefont {L.}~\bibnamefont {Di~Virgilio}}, \bibinfo
  {author} {\bibfnamefont {A.}~\bibnamefont {Greco}}, \bibinfo {author}
  {\bibfnamefont {M.}~\bibnamefont {Ceccanti}}, \bibinfo {author}
  {\bibfnamefont {E.}~\bibnamefont {Arimondo}}, \bibinfo {author}
  {\bibfnamefont {D.}~\bibnamefont {Ciampini}}, \bibinfo {author}
  {\bibfnamefont {I.~I.}\ \bibnamefont {Ryabtsev}}, \bibinfo {author}
  {\bibfnamefont {I.~I.}\ \bibnamefont {Beterov}}, \ and\ \bibinfo {author}
  {\bibfnamefont {O.}~\bibnamefont {Morsch}},\ }\href {\doibase
  10.1103/PhysRevA.100.030501} {\bibfield  {journal} {\bibinfo  {journal}
  {Phys. Rev. A}\ }\textbf {\bibinfo {volume} {100}},\ \bibinfo {pages}
  {030501} (\bibinfo {year} {2019}{\natexlab{a}})}\BibitemShut {NoStop}%
\bibitem [{\citenamefont {Tallant}\ \emph {et~al.}(2010)\citenamefont
  {Tallant}, \citenamefont {Booth},\ and\ \citenamefont
  {Shaffer}}]{PhysRevA.82.063406}%
  \BibitemOpen
  \bibfield  {author} {\bibinfo {author} {\bibfnamefont {J.}~\bibnamefont
  {Tallant}}, \bibinfo {author} {\bibfnamefont {D.}~\bibnamefont {Booth}}, \
  and\ \bibinfo {author} {\bibfnamefont {J.~P.}\ \bibnamefont {Shaffer}},\
  }\href {\doibase 10.1103/PhysRevA.82.063406} {\bibfield  {journal} {\bibinfo
  {journal} {Phys. Rev. A}\ }\textbf {\bibinfo {volume} {82}},\ \bibinfo
  {pages} {063406} (\bibinfo {year} {2010})}\BibitemShut {NoStop}%
\bibitem [{\citenamefont {Magnani}\ \emph {et~al.}(2020)\citenamefont
  {Magnani}, \citenamefont {Mojica-Casique},\ and\ \citenamefont
  {Marcassa}}]{Magnani2020}%
  \BibitemOpen
  \bibfield  {author} {\bibinfo {author} {\bibfnamefont {B.}~\bibnamefont
  {Magnani}}, \bibinfo {author} {\bibfnamefont {C.}~\bibnamefont
  {Mojica-Casique}}, \ and\ \bibinfo {author} {\bibfnamefont {L.~G.}\
  \bibnamefont {Marcassa}},\ }\href {\doibase 10.1088/1361-6455/ab6a34}
  {\bibfield  {journal} {\bibinfo  {journal} {Journal of Physics B: Atomic,
  Molecular and Optical Physics}\ }\textbf {\bibinfo {volume} {53}},\ \bibinfo
  {pages} {064004} (\bibinfo {year} {2020})}\BibitemShut {NoStop}%
\bibitem [{\citenamefont {Reiser}\ and\ \citenamefont
  {Sch\"achter}(2013)}]{PhysRevA.87.033801}%
  \BibitemOpen
  \bibfield  {author} {\bibinfo {author} {\bibfnamefont {A.}~\bibnamefont
  {Reiser}}\ and\ \bibinfo {author} {\bibfnamefont {L.}~\bibnamefont
  {Sch\"achter}},\ }\href {\doibase 10.1103/PhysRevA.87.033801} {\bibfield
  {journal} {\bibinfo  {journal} {Phys. Rev. A}\ }\textbf {\bibinfo {volume}
  {87}},\ \bibinfo {pages} {033801} (\bibinfo {year} {2013})}\BibitemShut
  {NoStop}%
\bibitem [{\citenamefont {Archimi}\ \emph
  {et~al.}(2019{\natexlab{b}})\citenamefont {Archimi}, \citenamefont
  {Simonelli}, \citenamefont {Di~Virgilio}, \citenamefont {Greco},
  \citenamefont {Ceccanti}, \citenamefont {Arimondo}, \citenamefont {Ciampini},
  \citenamefont {Ryabtsev}, \citenamefont {Beterov},\ and\ \citenamefont
  {Morsch}}]{PhysRevA.100.030501}%
  \BibitemOpen
  \bibfield  {author} {\bibinfo {author} {\bibfnamefont {M.}~\bibnamefont
  {Archimi}}, \bibinfo {author} {\bibfnamefont {C.}~\bibnamefont {Simonelli}},
  \bibinfo {author} {\bibfnamefont {L.}~\bibnamefont {Di~Virgilio}}, \bibinfo
  {author} {\bibfnamefont {A.}~\bibnamefont {Greco}}, \bibinfo {author}
  {\bibfnamefont {M.}~\bibnamefont {Ceccanti}}, \bibinfo {author}
  {\bibfnamefont {E.}~\bibnamefont {Arimondo}}, \bibinfo {author}
  {\bibfnamefont {D.}~\bibnamefont {Ciampini}}, \bibinfo {author}
  {\bibfnamefont {I.~I.}\ \bibnamefont {Ryabtsev}}, \bibinfo {author}
  {\bibfnamefont {I.~I.}\ \bibnamefont {Beterov}}, \ and\ \bibinfo {author}
  {\bibfnamefont {O.}~\bibnamefont {Morsch}},\ }\href {\doibase
  10.1103/PhysRevA.100.030501} {\bibfield  {journal} {\bibinfo  {journal}
  {Phys. Rev. A}\ }\textbf {\bibinfo {volume} {100}},\ \bibinfo {pages}
  {030501} (\bibinfo {year} {2019}{\natexlab{b}})}\BibitemShut {NoStop}%
\bibitem [{\citenamefont {Mack}\ \emph {et~al.}(2015)\citenamefont {Mack},
  \citenamefont {Grimmel}, \citenamefont {Karlewski}, \citenamefont
  {S\'ark\'any}, \citenamefont {Hattermann},\ and\ \citenamefont
  {Fort\'agh}}]{PhysRevA.92.012517}%
  \BibitemOpen
  \bibfield  {author} {\bibinfo {author} {\bibfnamefont {M.}~\bibnamefont
  {Mack}}, \bibinfo {author} {\bibfnamefont {J.}~\bibnamefont {Grimmel}},
  \bibinfo {author} {\bibfnamefont {F.}~\bibnamefont {Karlewski}}, \bibinfo
  {author} {\bibfnamefont {L.~m.~H.}\ \bibnamefont {S\'ark\'any}}, \bibinfo
  {author} {\bibfnamefont {H.}~\bibnamefont {Hattermann}}, \ and\ \bibinfo
  {author} {\bibfnamefont {J.}~\bibnamefont {Fort\'agh}},\ }\href {\doibase
  10.1103/PhysRevA.92.012517} {\bibfield  {journal} {\bibinfo  {journal} {Phys.
  Rev. A}\ }\textbf {\bibinfo {volume} {92}},\ \bibinfo {pages} {012517}
  (\bibinfo {year} {2015})}\BibitemShut {NoStop}%
\bibitem [{\citenamefont {van Ninhuijs}\ \emph {et~al.}(2021)\citenamefont {van
  Ninhuijs}, \citenamefont {Daamen}, \citenamefont {Beckers},\ and\
  \citenamefont {Luiten}}]{ninhuijs2021design}%
  \BibitemOpen
  \bibfield  {author} {\bibinfo {author} {\bibfnamefont {M.~A.~W.}\
  \bibnamefont {van Ninhuijs}}, \bibinfo {author} {\bibfnamefont {K.~A.}\
  \bibnamefont {Daamen}}, \bibinfo {author} {\bibfnamefont {J.}~\bibnamefont
  {Beckers}}, \ and\ \bibinfo {author} {\bibfnamefont {O.~J.}\ \bibnamefont
  {Luiten}},\ }\href {\doibase 10.1063/5.0037846} {\bibfield  {journal}
  {\bibinfo  {journal} {Rev. Sci. Instrum.}\ }\textbf {\bibinfo {volume}
  {92}},\ \bibinfo {pages} {013506} (\bibinfo {year} {2021})}\BibitemShut
  {NoStop}%
\bibitem [{\citenamefont {Covey}\ \emph {et~al.}(2019)\citenamefont {Covey},
  \citenamefont {Sipahigil},\ and\ \citenamefont {Saffman}}]{Covey2019}%
  \BibitemOpen
  \bibfield  {author} {\bibinfo {author} {\bibfnamefont {J.~P.}\ \bibnamefont
  {Covey}}, \bibinfo {author} {\bibfnamefont {A.}~\bibnamefont {Sipahigil}}, \
  and\ \bibinfo {author} {\bibfnamefont {M.}~\bibnamefont {Saffman}},\ }\href
  {\doibase 10.1103/PhysRevA.100.012307} {\bibfield  {journal} {\bibinfo
  {journal} {Phys. Rev. A}\ }\textbf {\bibinfo {volume} {100}},\ \bibinfo
  {pages} {012307} (\bibinfo {year} {2019})}\BibitemShut {NoStop}%
\bibitem [{\citenamefont {Han}\ \emph {et~al.}(2018)\citenamefont {Han},
  \citenamefont {Vogt}, \citenamefont {Gross}, \citenamefont {Jaksch},
  \citenamefont {Kiffner},\ and\ \citenamefont {Li}}]{PhysRevLett.120.093201}%
  \BibitemOpen
  \bibfield  {author} {\bibinfo {author} {\bibfnamefont {J.}~\bibnamefont
  {Han}}, \bibinfo {author} {\bibfnamefont {T.}~\bibnamefont {Vogt}}, \bibinfo
  {author} {\bibfnamefont {C.}~\bibnamefont {Gross}}, \bibinfo {author}
  {\bibfnamefont {D.}~\bibnamefont {Jaksch}}, \bibinfo {author} {\bibfnamefont
  {M.}~\bibnamefont {Kiffner}}, \ and\ \bibinfo {author} {\bibfnamefont
  {W.}~\bibnamefont {Li}},\ }\href {\doibase 10.1103/PhysRevLett.120.093201}
  {\bibfield  {journal} {\bibinfo  {journal} {Phys. Rev. Lett.}\ }\textbf
  {\bibinfo {volume} {120}},\ \bibinfo {pages} {093201} (\bibinfo {year}
  {2018})}\BibitemShut {NoStop}%
\bibitem [{\citenamefont {Kumar}\ \emph {et~al.}(2023)\citenamefont {Kumar},
  \citenamefont {Suleymanzade}, \citenamefont {Stone}, \citenamefont {Taneja},
  \citenamefont {Anferov}, \citenamefont {Schuster},\ and\ \citenamefont
  {Simon}}]{Kumar2023Quantum}%
  \BibitemOpen
  \bibfield  {author} {\bibinfo {author} {\bibfnamefont {A.}~\bibnamefont
  {Kumar}}, \bibinfo {author} {\bibfnamefont {A.}~\bibnamefont {Suleymanzade}},
  \bibinfo {author} {\bibfnamefont {M.}~\bibnamefont {Stone}}, \bibinfo
  {author} {\bibfnamefont {L.}~\bibnamefont {Taneja}}, \bibinfo {author}
  {\bibfnamefont {A.}~\bibnamefont {Anferov}}, \bibinfo {author} {\bibfnamefont
  {D.~I.}\ \bibnamefont {Schuster}}, \ and\ \bibinfo {author} {\bibfnamefont
  {J.}~\bibnamefont {Simon}},\ }\href {\doibase 10.1038/s41586-023-04567-8}
  {\bibfield  {journal} {\bibinfo  {journal} {Nature}\ }\textbf {\bibinfo
  {volume} {615}},\ \bibinfo {pages} {614} (\bibinfo {year}
  {2023})}\BibitemShut {NoStop}%
\bibitem [{\citenamefont {Bohorquez}\ \emph {et~al.}(2023)\citenamefont
  {Bohorquez}, \citenamefont {Chinnarasu}, \citenamefont {Isaacs},
  \citenamefont {Booth}, \citenamefont {Beck}, \citenamefont {McDermott},\ and\
  \citenamefont {Saffman}}]{PhysRevA.108.022805}%
  \BibitemOpen
  \bibfield  {author} {\bibinfo {author} {\bibfnamefont {J.~C.}\ \bibnamefont
  {Bohorquez}}, \bibinfo {author} {\bibfnamefont {R.}~\bibnamefont
  {Chinnarasu}}, \bibinfo {author} {\bibfnamefont {J.}~\bibnamefont {Isaacs}},
  \bibinfo {author} {\bibfnamefont {D.}~\bibnamefont {Booth}}, \bibinfo
  {author} {\bibfnamefont {M.}~\bibnamefont {Beck}}, \bibinfo {author}
  {\bibfnamefont {R.}~\bibnamefont {McDermott}}, \ and\ \bibinfo {author}
  {\bibfnamefont {M.}~\bibnamefont {Saffman}},\ }\href {\doibase
  10.1103/PhysRevA.108.022805} {\bibfield  {journal} {\bibinfo  {journal}
  {Phys. Rev. A}\ }\textbf {\bibinfo {volume} {108}},\ \bibinfo {pages}
  {022805} (\bibinfo {year} {2023})}\BibitemShut {NoStop}%
\bibitem [{\citenamefont {Wade}\ \emph {et~al.}(2017)\citenamefont {Wade},
  \citenamefont {Šibalić}, \citenamefont {de~Melo}, \citenamefont {Kondo},
  \citenamefont {Adams},\ and\ \citenamefont {Weatherill}}]{Wade2017}%
  \BibitemOpen
  \bibfield  {author} {\bibinfo {author} {\bibfnamefont {C.~G.}\ \bibnamefont
  {Wade}}, \bibinfo {author} {\bibfnamefont {N.}~\bibnamefont {Šibalić}},
  \bibinfo {author} {\bibfnamefont {N.~R.}\ \bibnamefont {de~Melo}}, \bibinfo
  {author} {\bibfnamefont {J.~M.}\ \bibnamefont {Kondo}}, \bibinfo {author}
  {\bibfnamefont {C.~S.}\ \bibnamefont {Adams}}, \ and\ \bibinfo {author}
  {\bibfnamefont {K.~J.}\ \bibnamefont {Weatherill}},\ }\href {\doibase
  10.1038/nphoton.2016.214} {\bibfield  {journal} {\bibinfo  {journal} {Nature
  Photonics}\ }\textbf {\bibinfo {volume} {11}},\ \bibinfo {pages} {40}
  (\bibinfo {year} {2017})}\BibitemShut {NoStop}%
\bibitem [{\citenamefont {Wade}\ \emph {et~al.}(2018)\citenamefont {Wade},
  \citenamefont {Marcuzzi}, \citenamefont {Levi}, \citenamefont {Kondo},
  \citenamefont {Lesanovsky}, \citenamefont {Adams},\ and\ \citenamefont
  {Weatherill}}]{Wade2018}%
  \BibitemOpen
  \bibfield  {author} {\bibinfo {author} {\bibfnamefont {C.~G.}\ \bibnamefont
  {Wade}}, \bibinfo {author} {\bibfnamefont {M.}~\bibnamefont {Marcuzzi}},
  \bibinfo {author} {\bibfnamefont {E.}~\bibnamefont {Levi}}, \bibinfo {author}
  {\bibfnamefont {J.~M.}\ \bibnamefont {Kondo}}, \bibinfo {author}
  {\bibfnamefont {I.}~\bibnamefont {Lesanovsky}}, \bibinfo {author}
  {\bibfnamefont {C.~S.}\ \bibnamefont {Adams}}, \ and\ \bibinfo {author}
  {\bibfnamefont {K.~J.}\ \bibnamefont {Weatherill}},\ }\href {\doibase
  10.1038/s41467-018-05597-4} {\bibfield  {journal} {\bibinfo  {journal}
  {Nature Communications}\ }\textbf {\bibinfo {volume} {9}},\ \bibinfo {pages}
  {3567} (\bibinfo {year} {2018})}\BibitemShut {NoStop}%
\bibitem [{\citenamefont {Ebadi}\ \emph {et~al.}(2022)\citenamefont {Ebadi},
  \citenamefont {Keesling}, \citenamefont {Cain}, \citenamefont {Wang},
  \citenamefont {Levine}, \citenamefont {Bluvstein}, \citenamefont {Semeghini},
  \citenamefont {Omran}, \citenamefont {Liu}, \citenamefont {Samajdar},
  \citenamefont {Luo}, \citenamefont {Nash}, \citenamefont {Gao}, \citenamefont
  {Barak}, \citenamefont {Farhi}, \citenamefont {Sachdev}, \citenamefont
  {Gemelke}, \citenamefont {Zhou}, \citenamefont {Choi}, \citenamefont
  {Pichler}, \citenamefont {Wang}, \citenamefont {Greiner}, \citenamefont
  {Vuletić},\ and\ \citenamefont {Lukin}}]{Sebadi2022}%
  \BibitemOpen
  \bibfield  {author} {\bibinfo {author} {\bibfnamefont {S.}~\bibnamefont
  {Ebadi}}, \bibinfo {author} {\bibfnamefont {A.}~\bibnamefont {Keesling}},
  \bibinfo {author} {\bibfnamefont {M.}~\bibnamefont {Cain}}, \bibinfo {author}
  {\bibfnamefont {T.~T.}\ \bibnamefont {Wang}}, \bibinfo {author}
  {\bibfnamefont {H.}~\bibnamefont {Levine}}, \bibinfo {author} {\bibfnamefont
  {D.}~\bibnamefont {Bluvstein}}, \bibinfo {author} {\bibfnamefont
  {G.}~\bibnamefont {Semeghini}}, \bibinfo {author} {\bibfnamefont
  {A.}~\bibnamefont {Omran}}, \bibinfo {author} {\bibfnamefont {J.-G.}\
  \bibnamefont {Liu}}, \bibinfo {author} {\bibfnamefont {R.}~\bibnamefont
  {Samajdar}}, \bibinfo {author} {\bibfnamefont {X.-Z.}\ \bibnamefont {Luo}},
  \bibinfo {author} {\bibfnamefont {B.}~\bibnamefont {Nash}}, \bibinfo {author}
  {\bibfnamefont {X.}~\bibnamefont {Gao}}, \bibinfo {author} {\bibfnamefont
  {B.}~\bibnamefont {Barak}}, \bibinfo {author} {\bibfnamefont
  {E.}~\bibnamefont {Farhi}}, \bibinfo {author} {\bibfnamefont
  {S.}~\bibnamefont {Sachdev}}, \bibinfo {author} {\bibfnamefont
  {N.}~\bibnamefont {Gemelke}}, \bibinfo {author} {\bibfnamefont
  {L.}~\bibnamefont {Zhou}}, \bibinfo {author} {\bibfnamefont {S.}~\bibnamefont
  {Choi}}, \bibinfo {author} {\bibfnamefont {H.}~\bibnamefont {Pichler}},
  \bibinfo {author} {\bibfnamefont {S.-T.}\ \bibnamefont {Wang}}, \bibinfo
  {author} {\bibfnamefont {M.}~\bibnamefont {Greiner}}, \bibinfo {author}
  {\bibfnamefont {V.}~\bibnamefont {Vuletić}}, \ and\ \bibinfo {author}
  {\bibfnamefont {M.~D.}\ \bibnamefont {Lukin}},\ }\href {\doibase
  10.1126/science.abo6587} {\bibfield  {journal} {\bibinfo  {journal}
  {Science}\ }\textbf {\bibinfo {volume} {376}},\ \bibinfo {pages} {1209}
  (\bibinfo {year} {2022})},\ \Eprint
  {http://arxiv.org/abs/https://www.science.org/doi/pdf/10.1126/science.abo6587}
  {https://www.science.org/doi/pdf/10.1126/science.abo6587} \BibitemShut
  {NoStop}%
\bibitem [{\citenamefont {Jau}\ \emph {et~al.}(2016)\citenamefont {Jau},
  \citenamefont {Hankin}, \citenamefont {Keating}, \citenamefont {Deutsch},\
  and\ \citenamefont {Biedermann}}]{Jau2016}%
  \BibitemOpen
  \bibfield  {author} {\bibinfo {author} {\bibfnamefont {Y.-Y.}\ \bibnamefont
  {Jau}}, \bibinfo {author} {\bibfnamefont {A.~M.}\ \bibnamefont {Hankin}},
  \bibinfo {author} {\bibfnamefont {T.}~\bibnamefont {Keating}}, \bibinfo
  {author} {\bibfnamefont {I.~H.}\ \bibnamefont {Deutsch}}, \ and\ \bibinfo
  {author} {\bibfnamefont {G.~W.}\ \bibnamefont {Biedermann}},\ }\href
  {\doibase 10.1038/nphys3487} {\bibfield  {journal} {\bibinfo  {journal}
  {Nature Physics}\ }\textbf {\bibinfo {volume} {12}},\ \bibinfo {pages} {71}
  (\bibinfo {year} {2016})}\BibitemShut {NoStop}%
\bibitem [{\citenamefont {Holloway}\ \emph {et~al.}(2022)\citenamefont
  {Holloway}, \citenamefont {Prajapati}, \citenamefont {Artusio-Glimpse},
  \citenamefont {Berweger}, \citenamefont {Simons}, \citenamefont {Kasahara},
  \citenamefont {Alù},\ and\ \citenamefont {Ziolkowski}}]{Hollowy2022}%
  \BibitemOpen
  \bibfield  {author} {\bibinfo {author} {\bibfnamefont {C.~L.}\ \bibnamefont
  {Holloway}}, \bibinfo {author} {\bibfnamefont {N.}~\bibnamefont {Prajapati}},
  \bibinfo {author} {\bibfnamefont {A.~B.}\ \bibnamefont {Artusio-Glimpse}},
  \bibinfo {author} {\bibfnamefont {S.}~\bibnamefont {Berweger}}, \bibinfo
  {author} {\bibfnamefont {M.~T.}\ \bibnamefont {Simons}}, \bibinfo {author}
  {\bibfnamefont {Y.}~\bibnamefont {Kasahara}}, \bibinfo {author}
  {\bibfnamefont {A.}~\bibnamefont {Alù}}, \ and\ \bibinfo {author}
  {\bibfnamefont {R.~W.}\ \bibnamefont {Ziolkowski}},\ }\href {\doibase
  10.1063/5.0088532} {\bibfield  {journal} {\bibinfo  {journal} {Applied
  Physics Letters}\ }\textbf {\bibinfo {volume} {120}},\ \bibinfo {pages}
  {204001} (\bibinfo {year} {2022})},\ \Eprint
  {http://arxiv.org/abs/https://pubs.aip.org/aip/apl/article-pdf/doi/10.1063/5.0088532/16449170/204001\_1\_online.pdf}
  {https://pubs.aip.org/aip/apl/article-pdf/doi/10.1063/5.0088532/16449170/204001\_1\_online.pdf}
  \BibitemShut {NoStop}%
\bibitem [{\citenamefont {Liu}\ \emph {et~al.}(2024)\citenamefont {Liu},
  \citenamefont {Zhang}, \citenamefont {Liu}, \citenamefont {Wang},
  \citenamefont {Ma}, \citenamefont {Han}, \citenamefont {Zhang}, \citenamefont
  {Shao}, \citenamefont {Zhang}, \citenamefont {Li}, \citenamefont {Chen},
  \citenamefont {Ding},\ and\ \citenamefont {Shi}}]{liu2024}%
  \BibitemOpen
  \bibfield  {author} {\bibinfo {author} {\bibfnamefont {B.}~\bibnamefont
  {Liu}}, \bibinfo {author} {\bibfnamefont {L.-H.}\ \bibnamefont {Zhang}},
  \bibinfo {author} {\bibfnamefont {Z.-K.}\ \bibnamefont {Liu}}, \bibinfo
  {author} {\bibfnamefont {Q.-F.}\ \bibnamefont {Wang}}, \bibinfo {author}
  {\bibfnamefont {Y.}~\bibnamefont {Ma}}, \bibinfo {author} {\bibfnamefont
  {T.-Y.}\ \bibnamefont {Han}}, \bibinfo {author} {\bibfnamefont {Z.-Y.}\
  \bibnamefont {Zhang}}, \bibinfo {author} {\bibfnamefont {S.-Y.}\ \bibnamefont
  {Shao}}, \bibinfo {author} {\bibfnamefont {J.}~\bibnamefont {Zhang}},
  \bibinfo {author} {\bibfnamefont {Q.}~\bibnamefont {Li}}, \bibinfo {author}
  {\bibfnamefont {H.-C.}\ \bibnamefont {Chen}}, \bibinfo {author}
  {\bibfnamefont {D.-S.}\ \bibnamefont {Ding}}, \ and\ \bibinfo {author}
  {\bibfnamefont {B.-S.}\ \bibnamefont {Shi}},\ }\href@noop {} {\enquote
  {\bibinfo {title} {Cavity-enhanced rydberg atom microwave receiver},}\ }
  (\bibinfo {year} {2024}),\ \Eprint {http://arxiv.org/abs/2404.06915}
  {arXiv:2404.06915 [physics.atom-ph]} \BibitemShut {NoStop}%
\bibitem [{\citenamefont {Simons}\ \emph {et~al.}(2019)\citenamefont {Simons},
  \citenamefont {Haddab}, \citenamefont {Gordon}, \citenamefont {Novotny},\
  and\ \citenamefont {Holloway}}]{simons2019embedding}%
  \BibitemOpen
  \bibfield  {author} {\bibinfo {author} {\bibfnamefont {M.~T.}\ \bibnamefont
  {Simons}}, \bibinfo {author} {\bibfnamefont {A.~H.}\ \bibnamefont {Haddab}},
  \bibinfo {author} {\bibfnamefont {J.~A.}\ \bibnamefont {Gordon}}, \bibinfo
  {author} {\bibfnamefont {D.}~\bibnamefont {Novotny}}, \ and\ \bibinfo
  {author} {\bibfnamefont {C.~L.}\ \bibnamefont {Holloway}},\ }\href {\doibase
  10.1109/ACCESS.2019.2949017} {\bibfield  {journal} {\bibinfo  {journal} {IEEE
  Access}\ }\textbf {\bibinfo {volume} {7}},\ \bibinfo {pages} {149936}
  (\bibinfo {year} {2019})},\ \bibinfo {note} {received August 31, 2019,
  accepted October 8, 2019, date of publication October 22, 2019, date of
  current version November 22, 2019}\BibitemShut {NoStop}%
\bibitem [{\citenamefont {Rodriguez~Fernandez}\ \emph
  {et~al.}(2024)\citenamefont {Rodriguez~Fernandez}, \citenamefont
  {Lefran~Torres}, \citenamefont {Cardoso}, \citenamefont {Kondo},
  \citenamefont {Saffman},\ and\ \citenamefont
  {Marcassa}}]{RodriguezFernandez2023}%
  \BibitemOpen
  \bibfield  {author} {\bibinfo {author} {\bibfnamefont {D.}~\bibnamefont
  {Rodriguez~Fernandez}}, \bibinfo {author} {\bibfnamefont {M.~A.}\
  \bibnamefont {Lefran~Torres}}, \bibinfo {author} {\bibfnamefont {M.~R.}\
  \bibnamefont {Cardoso}}, \bibinfo {author} {\bibfnamefont {J.~D.~M.}\
  \bibnamefont {Kondo}}, \bibinfo {author} {\bibfnamefont {M.}~\bibnamefont
  {Saffman}}, \ and\ \bibinfo {author} {\bibfnamefont {L.~G.}\ \bibnamefont
  {Marcassa}},\ }\href {\doibase 10.1007/s00340-024-08190-4} {\bibfield
  {journal} {\bibinfo  {journal} {Applied Physics B}\ }\textbf {\bibinfo
  {volume} {130}},\ \bibinfo {pages} {60} (\bibinfo {year} {2024})}\BibitemShut
  {NoStop}%
\bibitem [{sup(2024)}]{supp}%
  \BibitemOpen
  \href@noop {} {\enquote {\bibinfo {title} {See supplemental material at [url]
  for details of the experimental setup, fluorescence measurement and
  electrometry calibration.}}\ } (\bibinfo {year} {2024})\BibitemShut {NoStop}%
\bibitem [{\citenamefont {Li}\ \emph {et~al.}(2003)\citenamefont {Li},
  \citenamefont {Mourachko}, \citenamefont {Noel},\ and\ \citenamefont
  {Gallagher}}]{LiPRA2003}%
  \BibitemOpen
  \bibfield  {author} {\bibinfo {author} {\bibfnamefont {W.}~\bibnamefont
  {Li}}, \bibinfo {author} {\bibfnamefont {I.}~\bibnamefont {Mourachko}},
  \bibinfo {author} {\bibfnamefont {M.~W.}\ \bibnamefont {Noel}}, \ and\
  \bibinfo {author} {\bibfnamefont {T.~F.}\ \bibnamefont {Gallagher}},\ }\href
  {\doibase 10.1103/PhysRevA.67.052502} {\bibfield  {journal} {\bibinfo
  {journal} {Phys. Rev. A}\ }\textbf {\bibinfo {volume} {67}},\ \bibinfo
  {pages} {052502} (\bibinfo {year} {2003})}\BibitemShut {NoStop}%
\bibitem [{\citenamefont {Wang}\ \emph {et~al.}(2023)\citenamefont {Wang},
  \citenamefont {Hou}, \citenamefont {Lu}, \citenamefont {Chang}, \citenamefont
  {Hao}, \citenamefont {Su}, \citenamefont {Bai}, \citenamefont {He},\ and\
  \citenamefont {Wang}}]{Wang2023AT}%
  \BibitemOpen
  \bibfield  {author} {\bibinfo {author} {\bibfnamefont {X.}~\bibnamefont
  {Wang}}, \bibinfo {author} {\bibfnamefont {X.}~\bibnamefont {Hou}}, \bibinfo
  {author} {\bibfnamefont {F.}~\bibnamefont {Lu}}, \bibinfo {author}
  {\bibfnamefont {R.}~\bibnamefont {Chang}}, \bibinfo {author} {\bibfnamefont
  {L.}~\bibnamefont {Hao}}, \bibinfo {author} {\bibfnamefont {W.}~\bibnamefont
  {Su}}, \bibinfo {author} {\bibfnamefont {J.}~\bibnamefont {Bai}}, \bibinfo
  {author} {\bibfnamefont {J.}~\bibnamefont {He}}, \ and\ \bibinfo {author}
  {\bibfnamefont {J.}~\bibnamefont {Wang}},\ }\href {\doibase
  10.1063/5.0141479} {\bibfield  {journal} {\bibinfo  {journal} {AIP Advances}\
  }\textbf {\bibinfo {volume} {13}},\ \bibinfo {pages} {035126} (\bibinfo
  {year} {2023})},\ \Eprint
  {http://arxiv.org/abs/https://pubs.aip.org/aip/adv/article-pdf/doi/10.1063/5.0141479/16785789/035126\_1\_online.pdf}
  {https://pubs.aip.org/aip/adv/article-pdf/doi/10.1063/5.0141479/16785789/035126\_1\_online.pdf}
  \BibitemShut {NoStop}%
\bibitem [{\citenamefont {Gerry}\ and\ \citenamefont
  {Knight}(2004)}]{Gerry_Knight_2004}%
  \BibitemOpen
  \bibfield  {author} {\bibinfo {author} {\bibfnamefont {C.}~\bibnamefont
  {Gerry}}\ and\ \bibinfo {author} {\bibfnamefont {P.}~\bibnamefont {Knight}},\
  }\href@noop {} {\emph {\bibinfo {title} {Introductory Quantum Optics}}}\
  (\bibinfo  {publisher} {Cambridge University Press},\ \bibinfo {year}
  {2004})\BibitemShut {NoStop}%
\end{thebibliography}%

\end{document}